\documentclass[letter]{aa} 
\usepackage{graphicx,natbib}
\usepackage{txfonts}
\newcommand{\Msun}{$M_\odot$}
\newcommand{\msun}{\ensuremath{\mathit{M}_{\odot}}}   
\newcommand{\sh}{\mathrm{sh}}
\newcommand{\rel}{\mathrm{rel}}
\newcommand{\mi}{\mathrm{min}}
\newcommand{\sync}{\mathrm{sync}}
\newcommand{\SSA}{\mathrm{SSA}}
\newcommand{\kin}{\mathrm{ej}}
\newcommand{\ej}{\mathrm{ej}}

\newcommand{\mdot}{\ensuremath{\dot{M}}}                             
\newcommand{\vinf}{\ensuremath{v_{\infty}}}                          

\begin{document} 

   \authorrunning{Moriya, Groh, \& Meynet}
   \titlerunning{LBV SN Progenitors and Episodic SN Radio LC Modulations}

   \title{
Episodic modulations in supernova radio light curves from
luminous blue variable supernova progenitor models 
    }

   \subtitle{}

   \author{Takashi J. Moriya\inst{1,2,3},          
	  Jose H. Groh\inst{4}, \and
	  Georges Meynet\inst{4}
          }

   \institute{
Kavli Institute for the Physics and Mathematics of the Universe (WPI),
Todai Institutes for Advanced Study, University of Tokyo, Kashiwanoha
5-1-5, Kashiwa, Chiba 277-8583, Japan; \email{takashi.moriya@ipmu.jp}\and
Department of Astronomy, Graduate School of Science, University of
Tokyo, Hongo 7-3-1, Bunkyo, Tokyo, Japan \and
Research Center for the Early Universe, Graduate School of Science, University of Tokyo, Hongo 7-3-1, Bunkyo, Tokyo, Japan
         \and
	 Geneva Observatory, Geneva University, Chemin des Maillettes
              51, 1290 Sauverny, Switzerland \\
             }

   \date{}

 
  \abstract
   {Ideally, one would like to know which type of core-collapse supernovae (SNe) is produced by different 
 progenitors and the channels of stellar evolution leading to these progenitors. These links have to be
 very well known to use the observed frequency of different types of SN events for probing the star formation rate 
 and massive star evolution in different types of galaxies.}
   {We investigate the link between luminous blue variable (LBV) as SN
   progenitors and the appearance of episodic light curve modulations
   in the radio light curves of the SN event.}
  {  We use the 20 $M_\odot$  and 25 $M_\odot$ models with rotation at solar metallicity,
   part of an extended grid of stellar models computed by the Geneva team. At their pre-SN stage, these two models
   have recently been shown to have spectra similar to those of LBV
   stars and possibly explode as Type IIb SNe. Based on the wind properties before the explosion, we derive the
   density structure of their circumstellar medium. This structure is used as input for computing the SN radio light curve.}
   {We find that the 20 $M_\odot$ model shows radio light curves with
   episodic luminosity modulations, similar to those observed in some
   Type IIb SNe. This occurs because the evolution of the 20~\Msun\
   model terminates in a region of the HR diagram where radiative
   stellar winds present strong density variations, caused by the
   bistability limit. The 25 $M_\odot$ model, ending its evolution in a zone of the HR diagram where no change of the mass-loss rates is expected, presents no such modulations in its radio SN light curve.}
   {
Our results reinforce the link between SN progenitors and LBV stars.
We also confirm the existence of a physical mechanism for a single star to have
episodic radio light curve modulations.
In the case of the 25 $M_\odot$ progenitors, we do not obtain
   modulations in the radio light curve, but our models may miss some outbursting behavior in the late stages
   of massive stars.}


   \maketitle
%

\section{Introduction}
In the past decade,  supernova (SN) progenitor surveys using pre-explosion images
have revealed the properties of massive stars shortly before
the explosions (\citealt{smartt2009} for a review).
Some SN progenitors have been linked to luminous blue variables
(LBVs) \citep[e.g.,][]{kotak2006,smith2007,gal-yam2009,moriya2013,mauerhan2013}.
However, until recently, LBVs were theoretically considered to be at the transitional
phase to Wolf-Rayet stars and their core does not collapse at this stage
\citep[e.g.,][]{maeder2000,langer2012}.
Recently, \citet{groh2013} showed that the theoretical spectra of
the rotating 20 \Msun\ and 25 \Msun\ pre-SN models are similar to those of LBVs,
and that some SN progenitors can be at the LBV stage at the time
of the core-collapse explosion. These models have also been suggested to
explode as Type IIb SNe (SNe IIb) \citep{groh2013}.

The possibility of LBVs being SN progenitors has been originally suggested by
\citet[][KV06 hereafter]{kotak2006} through the interpretation of SN radio light curves (LCs).
Radio emission from SNe is caused by the interaction
between SN ejecta and the progenitors' circumstellar media (CSM). Therefore, 
the mass-loss history of the SN progenitor is imprinted in SN radio LCs
\citep[e.g.,][]{weiler2002,chevalier2006b,chevalier2006a}.
Radio emission from some SNe is known to have episodic luminosity enhancements, as has been
clearly observed in, e.g.,
SN IIb 2001ig \citep{ryder2004},
SN IIb 2003bg \citep{soderberg2006},
SN IIb 2008ax \citep{roming2009},
SN IIb 2011ei \citep{milisavljevic2013},
SN Ic 1998bw \citep{kulkarni1998}, and
SN IIL 1979C \citep{weiler1992}.
KV06 suggested that the timescales of the episodic radio modulations
are consistent with S Doradus-type mass loss, which occurs only in LBVs. During the S Doradus variability, the star crosses the bistability limit, 
in which the mass-loss rate (\mdot) and wind terminal velocity (\vinf) change abruptly \citep[e.g.][]{ghd09,ghd11,gv11}. The changes are regulated by 
Fe recombination in the inner wind, leading to changes in the wind driving and thus in \mdot\ and \vinf\ \citep{pauldrach90,Vink1999,Vink2002}.
A density change by a factor of $\sim 4-10$ would be expected, and this
recurrent behavior would create an inhomogeneous CSM around the LBV at
the pre-SN stage, causing the observed radio LC variations (KV06).

\begin{figure*}
\centering
\includegraphics[width=0.95\columnwidth]{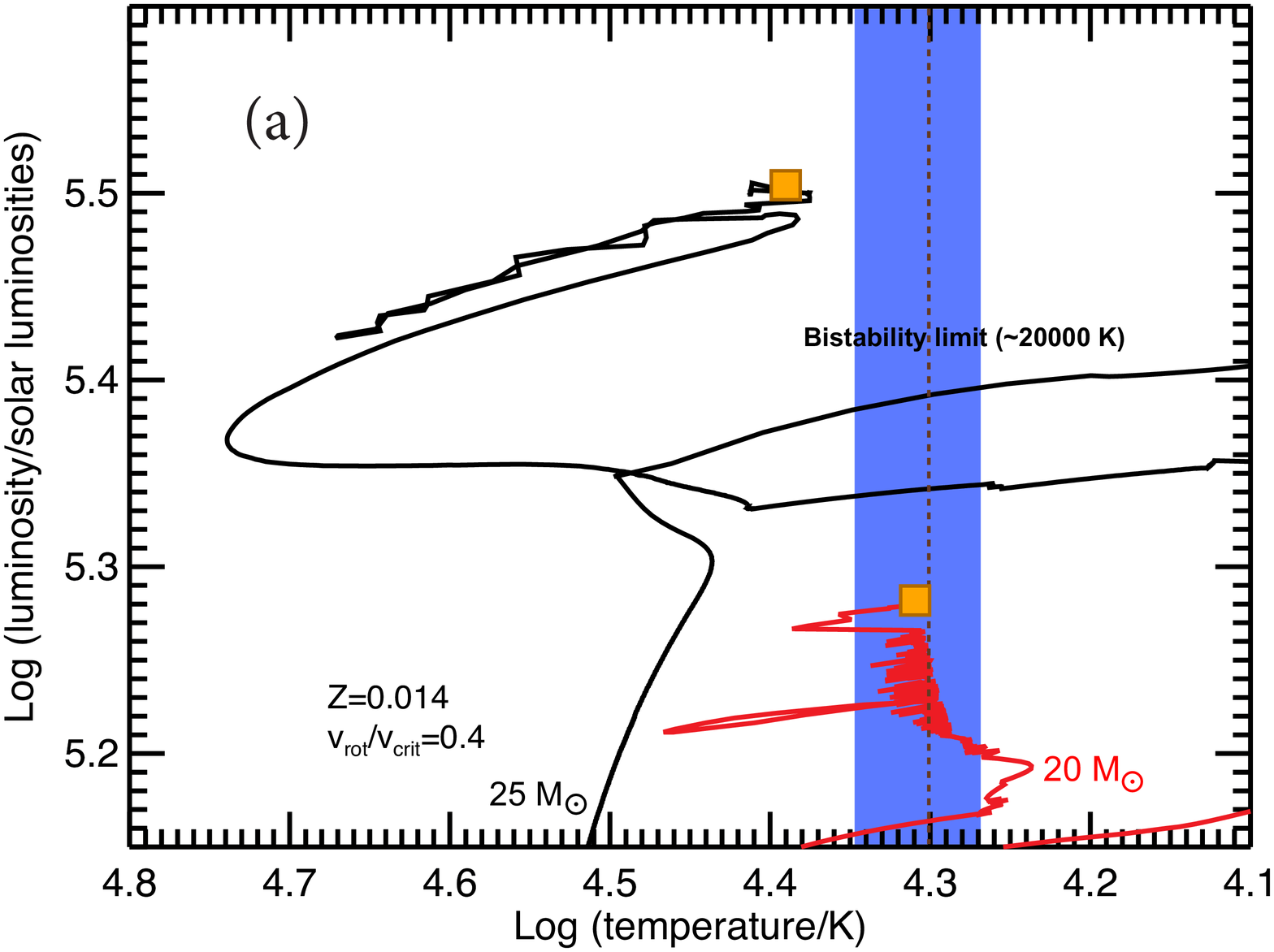}
\includegraphics[width=0.95\columnwidth]{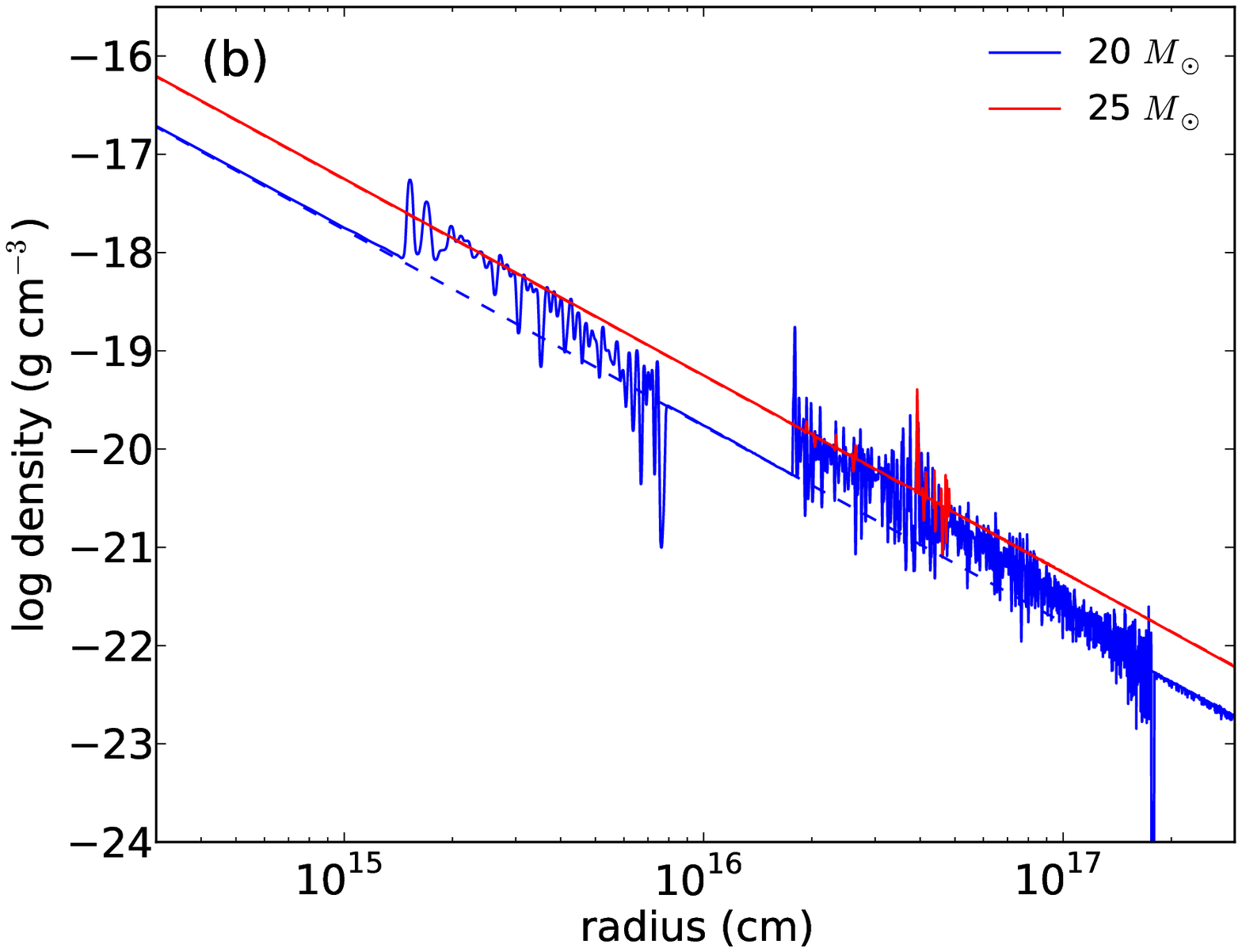}
\caption{
(a) End part of the evolutionary tracks in the HR diagram for the 20 and
 25 \msun\ rotating models. The end point is shown as an orange square. The bistability limit, where jumps in $\mdot$ and $\vinf$ occur, is indicated in blue. The vertical line corresponds to the bistability limit computed for the end point according to the \citet{Vink1999} recipe. (b)  CSM density structures obtained from the evolutionary models (solid lines).
The CSM densities with dashed lines are the CSM without the mass loss variations
and used to obtain the reference LCs.
}
\label{CSMdensity}
\end{figure*}

In this Letter, we investigate the behavior of the 20 and 25 \msun\
LBV models centuries before their explosions.
We show that the 20 \Msun\ SN progenitor model flirts with the
bistability limit, presenting variable mass loss and an inhomogeneous CSM
density structure at the time of the explosion. We present a
quantitative model for the SN radio LC that can naturally explain the
episodic radio LC modulations from the single-star evolutionary point of
view. Our results bring further support to the idea that a fraction of LBVs can be the end stage of massive stars.

\section{Stellar Evolution}

\subsection{Pre-Supernova Models at the LBV Phase}

The 20 and 25 \msun\ stellar models discussed here
have been computed by \citet{ekstrom2012}
for an initial metallicity $Z=0.014$. 
The physical ingredients used to compute these models can be found in this reference as well
as a presentation of their main characteristics. 
Let us just recall here a few points allowing to make the present paper self-explanatory.
The time averaged equatorial rotation velocities during the main-sequence
phase are equal to 217 (20 $M_\odot$) and 209 km s$^{-1}$ (25 $M_\odot$).
An overshooting equal to 10\% of the pressure scale at the border of the Schwarzschild convective core
has been accounted for. 
The radiative mass-loss rate adopted
is from \citet{Vink2001}. In the domains not covered by this
prescription, we use \citet{deJager1988}. 
We have applied a correction factor due to rotation to
the radiative mass-loss rate as described in \citet{maeder2000b}. For the two models discussed here, this correcting factor has very little
impact on the mass-loss rates.
In the red supergiant (RSG) phase, when some of the most external
layers of the stellar envelope exceed the Eddington
luminosity of the star, the mass-loss rate of the
star (computed according to the prescription described above)
is increased by a factor of 3.
As emphasized by \citet{ekstrom2012}, this prescription
gives \mdot\ during the RSG phase compatible with those of RSGs obtained by \citet{vanLoon2005}.

These two models end their nuclear lifetimes in the blue part of the
HR diagram (Fig. \ref{CSMdensity}a).
\citet{groh2013} showed that the spectrum of these stars
at the pre-SN stage looks like quiescent LBV stars, showing for the first time
that single star evolution may produce LBV-type progenitors of core-collapse SN events.
Interestingly, Fig. 1a shows that the 20 $M_\odot$ model ends its evolution when its
surface conditions are just at the frontier between two regimes of mass
loss that characterizes the bistability limit, as described above.
On the basis of mass-loss properties, \citet{Vink2002} proposed a
connection between the bistability limit and the LBV stars. 
Here we show from a stellar evolution point of view that indeed there may be such a connection.

From Fig. \ref{CSMdensity}a, we see that
our 20 $M_\odot$ model has the effective temperatures at the end of its evolution which
oscillate around $\log (T_\mathrm{eff}/\mathrm{K})=$ 4.3, implying variations of the mass-loss rates between about $1.2\times10^{-5}$ (on the hot side of the bistability limit)
and $15\times10^{-5}$ $M_\odot~\mathrm{yr^{-1}}$ (on the cool side). Thus this model would nicely fit the picture described in \citet{Vink1999,Vink2002}.
On the other hand, in the case of our 25 $M_\odot$ model, the end point is at
a too high effective temperature for such a process to occur.  Actually, both models could present another type
of mass-loss variability before they explode. For example, S Doradus
variability could still occur and produce variable $\mdot$ near the end
stage. Unfortunately, at the moment there is no accepted theory to explain the S
Doradus variability, so they cannot be self-consitently included in the evolutionary models.

We have thus two models presenting an LBV-type spectrum at the pre-SN stage. One, the 20 $M_\odot$
model,  just stops in the vicinity of the bistability limit while the other has its end point far from this limit.
Let us now study the consequences of these two types of behavior on the SN radio LCs.
For this purpose, we need first to derive the CSM properties of the pre-SN stars.

\subsection{Circumstellar Media}\label{sec:csm}
To construct the CSM density structures from the mass-loss rates and the
wind velocities obtained by the stellar evolutionary model,
we perform one-dimensional spherically-symmetric numerical hydrodynamics
calculations with \verb|ZEUS-MP2| version 2.1.2 \citep{hayes2006}. 
The CSM structures of
the region between $1.5\times 10^{13}$ cm ($6R_\star$, where $R_\star$
is the stellar radius of the 20 $M_\odot$ model)
and $3\times 10^{17}$ cm ($10^{5}R_\star$)
are followed by setting 
the inner boundary conditions at $1.5\times 10^{13}$
cm based on the mass-loss rates and wind velocities.

The CSM density structures obtained are shown in Fig. \ref{CSMdensity}b.
In the 20 \Msun\ model, there exist two extended high-density regions.
The two regions correspond to the two enhanced mass-loss periods 
caused by the star crossing to the cool side of the bistability limit.
The small-scale density variations are due to the rapid variations in the
mass-loss rate. On the contrary, the 25 \msun\ model does not have significant variations
in the mass-loss rate shortly before the core collapse and
it does not have any extended high-density regions in the CSM near the progenitor.

\section{Supernova Radio Light Curve}\label{sec:radioLC}

\subsection{Radio Light Curve Model}\label{LCmodel}
We calculate SN radio LCs based on the CSM density structures obtained in
Section \ref{sec:csm}.
SN radio emission is considered to be 
synchrotron emission from the accelerated electrons at the forward shock.
We obtain the synchrotron luminosity by following the formalism
developed by \citet{fransson1998,bjornsson2004} (see also \citealt{maeda2012}).
The synchrotron luminosity $L_\nu$ at a frequency $\nu$ is approximated as
\begin{equation}
\nu L_\nu \approx \pi R_\sh^2 V_\sh n_\rel 
\left(\frac{\gamma_\nu}{\gamma_\mi}\right)^{1-p}
\gamma_\nu m_e c^2 \left[1+\frac{t_\sync(\nu)}{t}\right]^{-1},
\label{radioluminosity}
\end{equation}
where $m_e$ is the electron mass and $c$ is the speed of light.
$R_\sh$ and $V_\sh$ are the radius and the velocity of the forward
shock, respectively.
$n_\rel$ is the number density of the relativistic electrons.
We set the distribution of the relativistic electrons as
$dn_\rel(\gamma)/d\gamma\propto \gamma^{-p}$ where $\gamma$ is the
Lorentz factor. We adopt $p=3$, which is the
canonical value for the synchrotron emission from SNe.
$\gamma_\mi$ is the minimum Lorentz factor of the accelerated electrons
and usually assumed to be $\gamma_\mi\sim 1$.
$\gamma_\nu=(2\pi m_e c\nu/eB)^{0.5}$, where $e$ is the electron charge
and $B$ is the magnetic field strength,
 is the Lorentz factor of electrons emitting at the
characteristic frequency $\nu$.
$t_\sync(\nu)=6\pi m_ec/\sigma_T\gamma_\nu B^2$ is
the cooling timescale of electrons emitting at $\nu$ by
the synchrotron cooling, where $\sigma_T$ is the Thomson cross section.
In addition, we take the synchrotron self-absorption (SSA) into account \citep[e.g.,][]{chevalier1998}.
Free-free absorption is ignored because of the low CSM density.
We use the SSA optical depth $\tau_\SSA=(\nu/\nu_\SSA)^{-(p+4)/2}$.
For $p=3$, the SSA frequency $\nu_\SSA$ is
$\approx 3\times 10^5(R_\sh\epsilon_e/\epsilon_B)^{2/7}B^{9/7}$ Hz
in cgs units,
where $\epsilon_e$ 
is the fraction of thermal energy in the shock used for the electron
acceleration
and $\epsilon_B$ is the fraction converted to the magnetic field energy.

We adopt the self-similar solution of \citet{chevalier1982}
for $R_\sh$ and $V_\sh$. 
The SN ejecta with kinetic energy $E_\kin$
and mass $M_\ej$ is assumed to have the density structure with the
two power-law components
($\rho\propto r^{-n}$ outside and $\rho\propto r^{-\delta}$ inside).
We adopt $n=10.2$ and $\delta=1.1$ which approximate the numerical
explosion of SN IIb/Ib/Ic progenitors \citep{matzner1999}.
For simplicity, we ignore the effect of the density jumps in CSM
on $R_\sh$ and $V_\sh$ as is also assumed in \citet{soderberg2006}.
We use the dashed density structures in Fig. \ref{CSMdensity}b
to obtain $R_\sh$ and $V_\sh$. 
We constrain $M_\ej$ by subtracting the remnant mass $1.4 M_\odot$ from
the progenitor mass obtained by \citet{ekstrom2012} and
$M_\ej=5.7~M_\odot$ ($20~M_\odot$ model) and $M_\ej=8.2 M_\odot$
($25 M_\odot$ model).

\begin{figure}
\centering
\includegraphics[width=0.95\columnwidth]{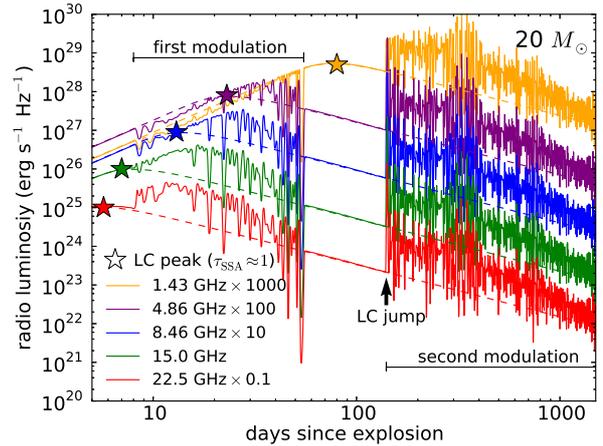}
\caption{
Synthesized radio LCs with the standard SN kinetic energy
 $E_\kin=10^{51}$ erg
from the 20 \Msun\ model (solid lines).
The reference LCs without the CSM density variations are also shown
(dashed lines, see Fig. \ref{CSMdensity}b). The star symbols indicate the LC
 peaks of the reference LCs.
}
\label{radioLC}
\end{figure}

\subsection{Results}

SN radio LCs obtained from the 20 \Msun\ LBV SN progenitor
exploded with the standard SN ejecta kinetic energy $E_\kin=10^{51} \mathrm{erg}$
are presented in Fig. \ref{radioLC}.
We assume $\epsilon_e=0.1$ and $\epsilon_B=0.1$ here.
The radio LCs show two episodic modulations.
They result from the two high-density regions due to the mass-loss
enhancements caused by the bistability limit shortly before the explosion.
The forward shock reaches the first high-density region
at around 8 days
since the explosion. At this time, the effect of the SSA is still
dominant at 1.43 GHz, 4.86 GHz, and 8.46 GHz
and the radio luminosities at these frequencies decrease due to the density
enhancement. 
As time passes, the SSA gets less effective and the radio luminosities
start to be enhanced by the density increase.
At the time when the forward shock reaches the second high-density region
at around 140 days,
the SSA is negligible in all the frequencies in Fig. \ref{radioLC}
and the radio luminosities are enhanced by about a factor $5$ on average in all the bands.
The radio LCs also have very short variations which are caused by the
small-scale density variations seen in Fig. \ref{CSMdensity}b.
However, these short-time variations should be smoothed by the
light-travel-time delays caused by the large emitting radii
 which are not taken into account in our modeling.
On the contrary, the 25 \Msun\ model does not show the LC
modulations as its surface temperature is far from from the temperature
of the bistability  limit
and there is no significant mass-loss increase shortly before the explosion. 

To detect the overall features of the radio LCs caused
by the LBV progenitor predicted here in Fig. \ref{radioLC},
we need to be sensitive to the radio luminosity of
$\sim 10^{24}\ \mathrm{erg~s^{-1}~Hz^{-1}}$.
Using the Extended Very Large Array, which is sensitive down to
$\sim 1 \mathrm{\mu Jy}$ \citep{perley2011},
the predicted SN radio LC features
can be observed if the corresponding SN appears within 30 Mpc.

\begin{figure*}
\centering
\includegraphics[width=0.95\columnwidth]{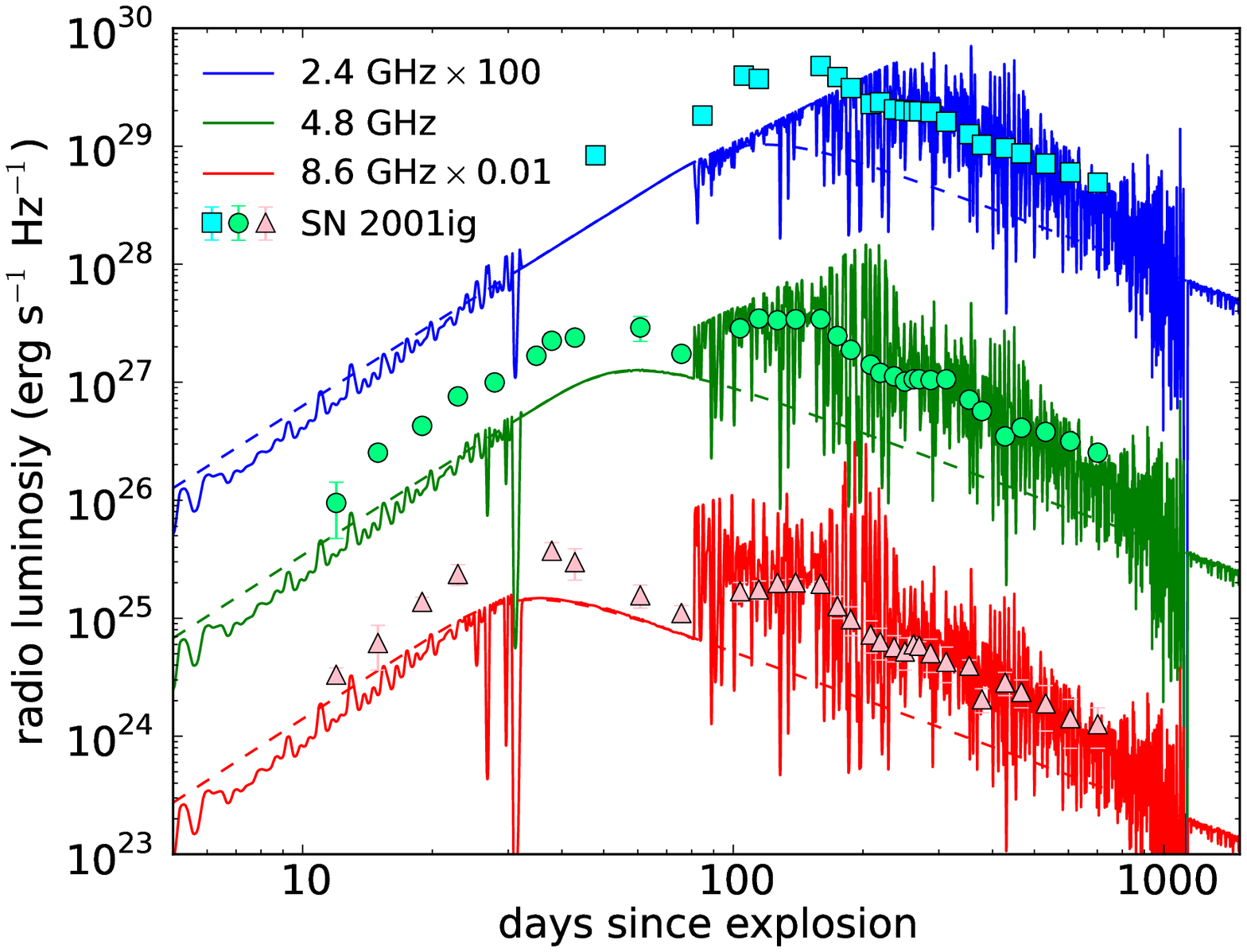}
\includegraphics[width=0.95\columnwidth]{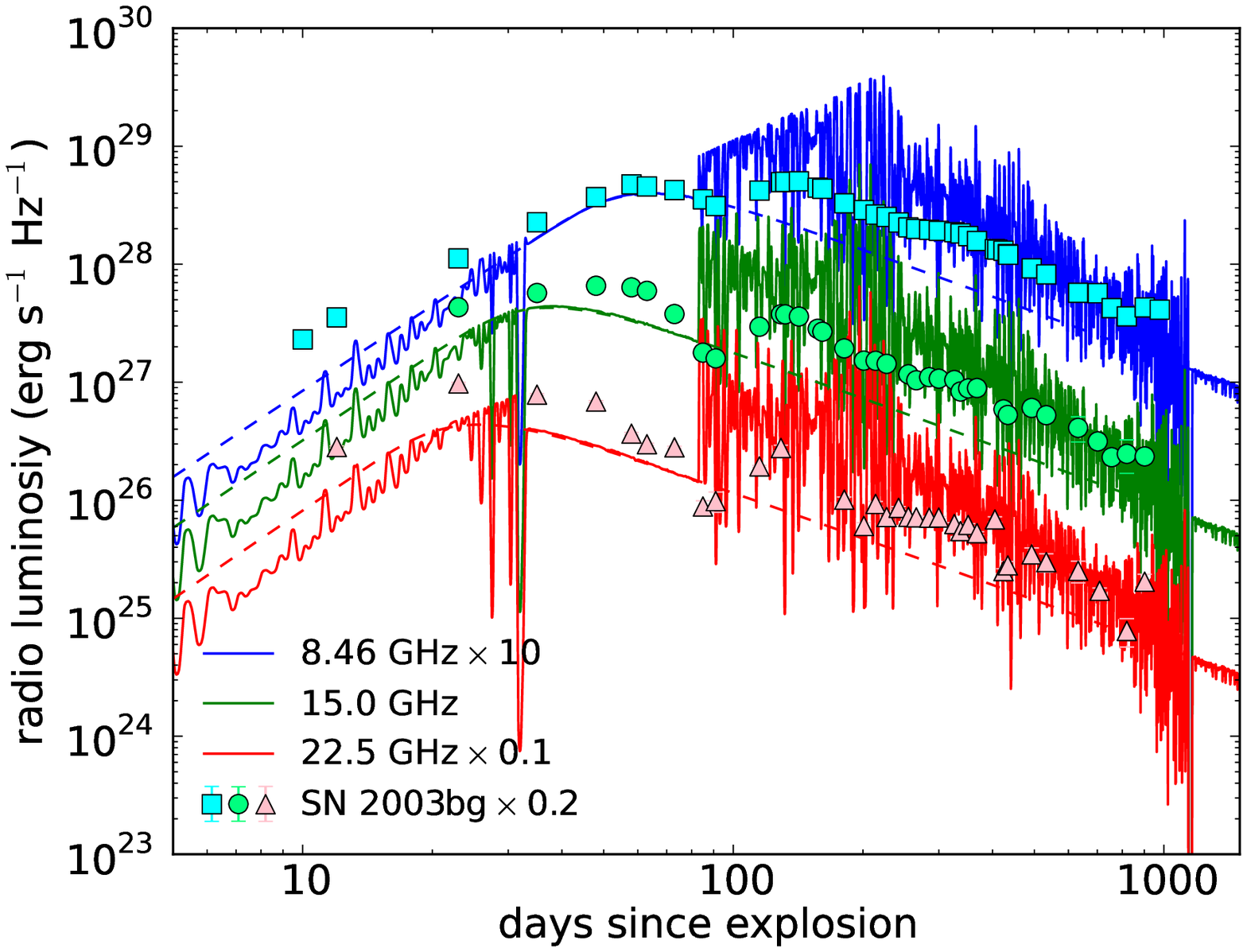}
\caption{
Comparisons between the model radio LCs from the 20 \Msun\ model
and the observed radio LCs of SNe IIb 2001ig \citep{ryder2004} and
2003bg \citep{soderberg2006}.
The frequencies of the observed radio LCs are 2.4 GHz (square),
4.8 GHz (circle), and 8.6 GHz (triangle) in the left panel.
The LCs of 2.4 GHz and 8.6 GHz are multiplied by 100 and 0.01,
 respectively, for the illustrative purpose.
In the right panel,
the frequencies of the observed radio LCs are 8.46 GHz (square),
15.0 GHz (circle), and 22.5 GHz (triangle).
The LCs of 8.46 GHz and 22.5 GHz are multiplied by 10 and 0.1,
respectively. The radio LCs of SN 2003bg are multiplied by 0.2 
 additionally to match the synthesized LCs.
}
\label{sn2003bg}
\end{figure*}

\section{Comparison with Observations}

The inhomogeneous
density structure around our model progenitors is reminiscent of that
qualitatively proposed by KV06 to explain the episodic
modulations in the radio LCs of some SNe. Let us now compare
our models with the observations.
We compare our 20 \Msun\ model with SNe IIb 2001ig \citep{ryder2004} and 2003bg
\citep{soderberg2006}. Their radio luminosities are higher than those we
obtained in the previous section. The radio luminosities
(Eq. \ref{radioluminosity}) follow
\begin{equation}
L_\nu\propto
 \epsilon_e\epsilon_B^{\frac{p+1}{4}}E_\kin^{\frac{3(n-3)}{2(n-2)}}
M_\ej^{-\frac{3(n-5)}{2(n-2)}}
\left(\frac{\dot{M}}{\vinf}\right)^{\frac{p+5}{4}-\frac{3}{n-2}}
\simeq
\epsilon_e\epsilon_B E_\kin^{1.3}M_\ej^{-0.95}
\left(\frac{\dot{M}}{\vinf}\right)^{1.6}.\label{dependence}
\end{equation}
The right-hand side of Eq. \ref{dependence} is for $p=3$ and $n=10.2$.
In this section, we show that the observed SN IIb radio LCs can be
explained by our model just by slightly changing the SN and predicted CSM
properties.

In Fig. \ref{sn2003bg}, we show the results of the comparisons.
On the left panel, our radio LC model is compared with SN IIb 2001ig.
To match the observed LCs, the CSM density is increased by a factor 3 
and we set $E_\ej=4\times 10^{51}$ erg. In addition, to adjust the time
of the LC peak before the second LC modulation, we set $\epsilon_e=0.2$
and $\epsilon_B=0.08$ (see $\tau_\SSA$ in Section \ref{LCmodel}).
We can see that the radio LC features of SN
2001ig are reproduced well by the above parameters which are not
so different from those of the standard model we presented in Section \ref{sec:radioLC}.
The episodic radio LC jump observed at around 100 days matches the
epoch when the model LCs start to
show the second modulation. This shows that the
changes in the CSM density that occur because of the crossing of the
bistability limit are able to explain
the modulation in the SN radio LC. The amount of the observed radio luminosity
increase 
also matches that in our model. This quantitative finding reinforces the link between LBVs as SN progenitors and the modulations in their radio LCs. In addition, it strengthens the idea that part of the SN IIb progenitors can be LBVs \citep[KV06,][]{groh2013}.
We confirm the existence of a physical mechanism for a single star to
make the inhomogeneous CSM around the SN 2001ig progenitor which has
often been related to the binary evolution \citep{ryder2004,ryder2006}.

The right panel shows the comparison with SN 2003bg.
The CSM density is increased by a factor 8 and $E_\kin=5\times 10^{51}$ erg in
the model shown. We keep $\epsilon_e=0.2$ and $\epsilon_B=0.08$.
The overall features of the observed LCs are reproduced.
However, the absolute luminosities of SN 2003bg are still higher
than those of the model, so the observations are scaled to match
the model. We can increase the luminosity by increasing $E_\kin$
but this makes the time of the LC jump much earlier than the observed time.
This indicates that there is a diversity in
the time of the mass-loss variations due
to the bistability limit, which is naturally expected.

The currently best observed SN IIb in radio is SN 1993J
\citep[e.g.,][]{weiler2007}. It does not show the episodic LC modulations we
present here but the progenitor is suggested to have had a sharp increase in
the mass-loss rate at around $10^4$ years before the explosion
\citep{weiler2007}.
Therefore, the timing of the mass-loss variation may be crucial for
observing episodic modulations in SN radio LCs.

We also note that the epochs of the first episodic luminosity
decrease/increase in the model
was not covered by the observations well. Observing these early modulations has the potential to probe mass loss immediately before the SN explosion, so early radio observations of SNe would be extremely invaluable.

\begin{acknowledgements}
We thank the referee for valuable comments.
T.J.M. thanks Keiichi Maeda for useful discussion.
Numerical computations were in part carried out on the general-purpose PC farm at Center for Computational Astrophysics, National Astronomical Observatory of Japan.
T.J.M. is supported by the Japan Society for the Promotion of Science
 Research Fellowship for Young Scientists (23$\cdot$5929).
This work is also supported by World Premier International Research Center Initiative (WPI Initiative), MEXT, Japan.
J.H.G. is supported by an Ambizione Fellowship of the Swiss National Science Foundation.
\end{acknowledgements}


\begin{thebibliography}{}

\bibitem[Bj{\"o}rnsson 
\& Fransson(2004)]{bjornsson2004} Bj{\"o}rnsson, C.-I., \& Fransson, C.\ 2004, \apj, 605, 823 

\bibitem[Chevalier(1998)]{chevalier1998} Chevalier, R.~A.\ 1998, 
\apj, 499, 810 

\bibitem[Chevalier(1982)]{chevalier1982} Chevalier, R.~A.\ 1982, 
\apj, 258, 790 

\bibitem[Chevalier 
\& Fransson(2006)]{chevalier2006a} Chevalier, R.~A., \& Fransson, C.\ 2006, \apj, 651, 381 

\bibitem[Chevalier et al.(2006)]{chevalier2006b} Chevalier, R.~A., 
Fransson, C., \& Nymark, T.~K.\ 2006, \apj, 641, 1029

\bibitem[de Jager et 
al.(1988)]{deJager1988} de Jager, C., Nieuwenhuijzen, H., \& van der Hucht, K.~A.\ 1988, \aaps, 72, 259 

\bibitem[Ekstr{\"o}m et 
al.(2012)]{ekstrom2012} Ekstr{\"o}m, S., Georgy, C., Eggenberger, P., et al.\ 2012, \aap, 537, A146

\bibitem[Gal-Yam 
\& Leonard(2009)]{gal-yam2009} Gal-Yam, A., \& Leonard, D.~C.\ 2009, \nat, 458, 865 

\bibitem[{{Groh} {et~al.}(2009){Groh}, {Hillier}, {Damineli}, {Whitelock},
  {Marang}, \& {Rossi}}]{ghd09}
{Groh}, J.~H., {Hillier}, D.~J., {Damineli}, A., {et~al.} 2009, \apj, 698, 1698

\bibitem[{{Groh} {et~al.}(2011){Groh}, {Hillier}, \& {Damineli}}]{ghd11}
{Groh}, J.~H., {Hillier}, D.~J., \& {Damineli}, A. 2011, \apj, 736, 46

\bibitem[{{Groh} \& {Vink}(2011)}]{gv11}
{Groh}, J.~H. \& {Vink}, J.~S. 2011, \aap, 531, L10

\bibitem[Groh et 
al.(2013)]{groh2013} Groh, J.~H., Meynet, G., \& Ekstr{\"o}m, S.\ 2013, \aap, 550, L7 

\bibitem[Fransson 
\& Bj{\"o}rnsson(1998)]{fransson1998} Fransson, C., \& Bj{\"o}rnsson, C.-I.\ 1998, \apj, 509, 861 

\bibitem[Hayes et al.(2006)]{hayes2006} Hayes, J.~C., Norman, 
M.~L., Fiedler, R.~A., et al.\ 2006, \apjs, 165, 188 


\bibitem[Kotak 
\& Vink(2006)]{kotak2006} Kotak, R., \& Vink, J.~S.\ 2006, \aap, 460,
	      L5 (KV06)

\bibitem[Kulkarni et al.(1998)]{kulkarni1998} Kulkarni, S.~R., 
Frail, D.~A., Wieringa, M.~H., et al.\ 1998, \nat, 395, 663 


\bibitem[Langer(2012)]{langer2012} Langer, N.\ 2012, \araa, 50, 107 

\bibitem[Maeda(2012)]{maeda2012} Maeda, K.\ 2012, \apj, 758, 81 

\bibitem[Maeder 
\& Meynet(2000a)]{maeder2000} Maeder, A., \& Meynet, G.\ 2000a, \araa, 38, 143 

\bibitem[Maeder 
\& Meynet(2000b)]{maeder2000b} Maeder, A., \& Meynet, G.\ 2000b, \aap, 361, 159 

\bibitem[Matzner 
\& McKee(1999)]{matzner1999} Matzner, C.~D., \& McKee, C.~F.\ 1999, \apj, 510, 379 

\bibitem[Mauerhan et al.(2013)]{mauerhan2013} Mauerhan, J.~C., 
Smith, N., Filippenko, A.~V., et al.\ 2013, \mnras, 430, 1801 

\bibitem[Milisavljevic et al.(2013)]{milisavljevic2013} Milisavljevic, 
D., Margutti, R., Soderberg, A.~M., et al.\ 2013, \apj, 767, 71 

\bibitem[Moriya et al.(2013)]{moriya2013} Moriya, T.~J., 
Blinnikov, S.~I., Tominaga, N., et al.\ 2013, \mnras, 428, 1020 

\bibitem[{{Pauldrach} \& {Puls}(1990)}]{pauldrach90}
{Pauldrach}, A.~W.~A. \& {Puls}, J. 1990, \aap, 237, 409

\bibitem[Perley et al.(2011)]{perley2011} Perley, R.~A., Chandler, 
C.~J., Butler, B.~J., \& Wrobel, J.~M.\ 2011, \apjl, 739, L1 

\bibitem[Roming et al.(2009)]{roming2009} Roming, P.~W.~A., 
Pritchard, T.~A., Brown, P.~J., et al.\ 2009, \apjl, 704, L118 

\bibitem[Ryder et al.(2006)]{ryder2006} Ryder, S.~D., Murrowood, 
C.~E., \& Stathakis, R.~A.\ 2006, \mnras, 369, L32 

\bibitem[Ryder et al.(2004)]{ryder2004} Ryder, S.~D., Sadler, 
E.~M., Subrahmanyan, R., et al.\ 2004, \mnras, 349, 1093

\bibitem[Smartt(2009)]{smartt2009} Smartt, S.~J.\ 2009, \araa, 47, 63 


\bibitem[{{Smith} {et~al.}(2007){Smith}, {Li}, {Foley}, {Wheeler}, {Pooley},
  {Chornock}, {Filippenko}, {Silverman}, {Quimby}, {Bloom}, \&
  {Hansen}}]{smith2007}
{Smith}, N., {Li}, W., {Foley}, R.~J., {et~al.} 2007, \apj, 666, 1116

\bibitem[Soderberg et al.(2006)]{soderberg2006} Soderberg, A.~M., 
Chevalier, R.~A., Kulkarni, S.~R., \& Frail, D.~A.\ 2006, \apj, 651, 1005 

\bibitem[van Loon et 
al.(2005)]{vanLoon2005} van Loon, J.~T., Marshall, J.~R., \& Zijlstra, A.~A.\ 2005, \aap, 442, 597 


\bibitem[Vink et 
al.(1999)]{Vink1999} Vink, J.~S., de Koter, A., \& Lamers, H.~J.~G.~L.~M.\ 1999, \aap, 350, 181 

\bibitem[Vink et 
al.(2001)]{Vink2001} Vink, J.~S., de Koter, A., \& Lamers, H.~J.~G.~L.~M.\ 2001, \aap, 369, 574 

\bibitem[{{Vink} \& {de Koter}(2002)}]{Vink2002}
{Vink}, J.~S. \& {de Koter}, A. 2002, \aap, 393, 543

\bibitem[Weiler et 
al.(2002)]{weiler2002} Weiler, K.~W., Panagia, N., Montes, M.~J., \& Sramek, R.~A.\ 2002, \araa, 40, 387 

\bibitem[Weiler et al.(1992)]{weiler1992} Weiler, K.~W., van Dyk, 
S.~D., Pringle, J.~E., \& Panagia, N.\ 1992, \apj, 399, 672 

\bibitem[Weiler et al.(2007)]{weiler2007} Weiler, K.~W., Williams, 
C.~L., Panagia, N., et al.\ 2007, \apj, 671, 1959 

\end{thebibliography}
\end{document}